\documentclass[aps,prl,twocolumn,superscriptaddress,showpacs]{revtex4}

\usepackage{graphicx} 

\begin{document}
\title{Superconducting Order Parameter
for the Even-denominator Fractional Quantum Hall Effect}

\author{Hantao Lu}
\affiliation{School of Physics, Korea Institute for Advanced
Study, Seoul 130-722, Korea}

\author{S. Das Sarma}
\affiliation{Condensed Matter Theory Center, Department of
Physics, University of Maryland, College Park, MD 20742}

\author{Kwon Park}
\affiliation{School of Physics, Korea Institute for Advanced
Study, Seoul 130-722, Korea}

\date{\today}

\begin{abstract}

\end{abstract}


\maketitle

{\bf One of the most intriguing phenomena in nature is the
fractional quantum Hall effect (FQHE) observed in the half-filled
second Landau level~\cite{Willett} which, arising in
even-denominator filling factors, $\nu=5/2$ and $7/2$, is completely
different from other FQHEs in its origin, all of which,
except for those two filling factors,
occur in odd-denominator fractions. Usually
formulated in terms of a trial wave function called the Moore-Read
Pfaffian wave function~\cite{MooreRead}, current leading
theories~\cite{GWW,PMBJ,SPJ_Nature,RezayiHaldane,ReadGreen}
attribute the origin of the 5/2 FQHE to the formation of Cooper pairs,
not of electron, but of the true quasi-particle of the system known as
composite fermion~\cite{Jain}. The
nature of superconductivity resulting from such Cooper pairing is particularly
puzzling in the sense that it apparently coexists with
strong magnetic fields, which poses an interesting dilemma since
the Meissner effect is {\it the} most important defining property
of superconductivity.
This apparent dilemma is resolved by the fact that composite fermions
do not respond to external magnetic field at even-denominator filling factors.
To provide direct evidence that it is composite fermions that actually form
the superconducting condensate, here, we develop a numerically exact method of
creating a Cooper pair of composite fermions and explicitly compute the superconducting
order parameter as a function of real space coordinates.
As results, in addition to direct evidence for
superconductivity, we obtain quantitative predictions for
superconducting coherence length. Obtaining such theoretical
predictions can serve as an important step toward fault-tolerant
topological quantum computation~\cite{DSFN,TQC_review}.}


We begin by emphasizing that no assumption is made throughout this
work regarding the nature of the 5/2 ground state. Our point of
view is that, if it exists, the superconducting order parameter
should arise naturally when computed via exact diagonalization.
Computing the superconducting order parameter, however, explicitly
depends on our choice of the quasi-particle in the ground state.
In the usual Fermi liquid, the quasi-particle is identified with
something very close to the free electron so that the
superconducting order parameter is computed in terms of the
``free'' electron operator. The situation is less clear for
strongly correlated electron systems. It is, thus, useful to
provide independent, unbiased evidence for superconductivity that
is not susceptible to our knowledge of the quasi-particle nature.
Such evidence is obtained in the ground state energy itself. Since
a prerequisite for superconductivity is the pairing between
quasi-particles, the ground state energy must oscillate depending
on whether the particle number is even or odd.



In order to obtain the FQHE ground state energy, it is convenient
to use a compact geometry with periodic boundary condition. In
this study, we use the so-called spherical geometry where
electrons are confined on the surface of a sphere which has a
magnetic monopole at its center~\cite{WuYang,Haldane}. The total
magnetic flux piercing through the spherical surface is given by
$2Q\phi_0$ with $\phi_0$, called the flux quantum, being equal to
$2\pi\hbar c/e$ and with $Q$, called the monopole strength, being
equal to either an integer or a half integer. Angular momentum
eigenstates in this geometry are known as the monopole harmonics
which form a basis set to span the single-particle Hilbert space
of each Landau level. Many-particle FQHE ground states are
obtained by numerically diagonalizing the inter-particle
interaction matrix within specific Landau levels; for example, the
diagonalization is done within the half-filled second Landau level
for the 5/2 FQHE. It is important to note that, due to the nature
of the monopole harmonics, the ratio between the number of
particles, $N$, and that of flux quanta, $2Q$, is not exactly
equal to the thermodynamic filling factor, $\nu$, in finite-size
systems; $N/2Q$ approaches $\nu$ as $N \rightarrow \infty$. It is
known from various studies that the 5/2 FQHE state occurs at
$2Q=2N-3$ with an apparent ``flux shift'' of
$-3$~\cite{GWW,PMBJ,SPJ_Nature,RezayiHaldane}.

\begin{figure}[t]
\begin{center}
\mbox{\includegraphics[width=3in]{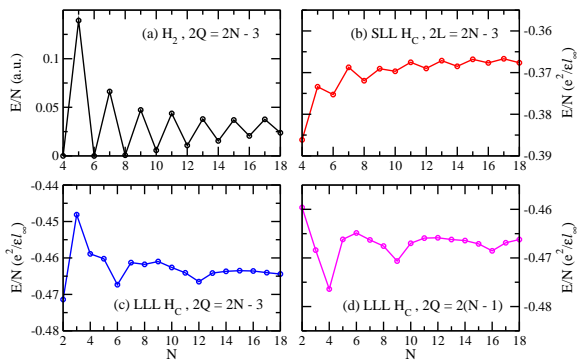}}
\end{center}
\caption{ {\bf Even-odd effect} The ground state energies per
particle are computed for the Pfaffian two-body interaction,
$H_2$, [top left panel], the Coulomb interaction, $H_C$, projected
in the second-Landau-level (SLL) [top right panel] and the
lowest-Landau-level (LLL) [bottom two panels]. For the Coulomb
interaction, the energy is given in units of $e^2/\varepsilon
l_{\infty}$ (where $l_{\infty}=l_B \sqrt{2\nu Q/N}$ with
$l_B=\sqrt{\hbar c/ eB}$~\cite{MdADS}) and the usual charge
background correction is taken into account. In the case of the
LLL $H_C$, two different sectors of flux quantum number, $2Q$, are
studied for a given particle number, $N$: the Pfaffian flux of
$2Q=2N-3$ and the composite-fermion-sea flux of $2Q=2N-2$.
}\label{fig1}
\end{figure}
Figure~\ref{fig1} shows the results of the ground state energy per
particle for the Coulomb interaction, $H_C$, of the half-filled
second-Landau-level (SLL) in comparison with that of the
half-filled lowest-Landau-level (LLL) as well as that of a model
two-body interaction, $H_2$, for which the Moore-Read Pfaffian
wave function is known to be very close to the exact ground
state~\cite{PPDS}. As one can see, even-odd effects are clearly
visible for both SLL $H_C$ and $H_2$ (top panels of
Fig.~\ref{fig1}), while the same is not true for LLL $H_C$ (bottom
panels). In fact, the behavior of the LLL $H_C$ ground state
energy is quite different from the SLL counterpart, showing a
peculiar pattern of repeating local minima at $N=4$, $9$, and $16$
(bottom right panel of Fig.~\ref{fig1}). The energy minima would
repeatedly appear whenever $N$ becomes $n^2$ with $n$ being a positive
integer if numerical diagonalization is possible for bigger
systems.

This behavior can be understood as follows. It is well established
by now that the ground state of the half-filled lowest Landau
level is the composite fermion (CF) sea which, while being a
compressible state in the thermodynamic limit, has a shell
structure in terms of the CF Landau levels in the finite-size
spherical geometry. The ground state energy exhibits local minima
whenever energy shells below the chemical potential become
completely filled, which occurs at $N=n^2$ with $n$ being a positive
integer~\cite{RR,MdA}. Note that the correct flux sector for the CF sea is
obtained at $2Q=2N-2$ for reasons provided by the CF theory. One
may wonder what happens if the same flux sector studied for SLL
$H_C$, i.~e., $2Q=2N-3$, is investigated in the LLL. The LLL
results are plotted in the bottom left panel of Fig.~\ref{fig1},
where it is shown that energy minimum positions are a little bit
different from those of the CF sea flux sector, while still not
exhibiting the even-odd effect. Here, energy minima occur for a combination of two reasons.
Now that the flux is shifted from that of the CF
sea, composite fermions are subject to small, residual field,
resulting in two ways of minimizing the total energy. The first is to
minimize the kinetic energy of composite fermions by filling the similar
shell structure as before, but with a different particle number sequence
of $N=n(n+1)$ with $n$ being a positive integer. The second is to
reduce any residual interaction energy by forming Wigner crystals of composite fermions,
which occur at $N=4$, 6, 8, 12, and 20 with $N$ corresponding to the number of
vertices in regular polyhedrons of Euclid. The energy reduction due to
the second effect is seen via a weak local minimum at $N=8$ and a kink at $N=4$.
The conclusion is that the SLL $H_C$ definitely exhibits the
even-odd effect, which strongly supports the existence of
superconductivity in a complete analogy with what happens in small
superconducting grains~\cite{Tinkham}. On the other hand, there is
no such evidence for the lowest Landau level in any flux sectors.
The clear signature of the even-odd effect in the half-filled SLL,
but not in the LLL, is one of the most important results in this
work. It is worthwhile to mention that the difference between the
second and lowest Landau-level physics stems from very subtle
quantitative changes in the Coulomb matrix elements parameterized
by the Haldane pseudopotential.

Emboldened by strong support from the even-odd effect, we now set
out to address the main issue in this paper: what are the
quasi-particles that are being paired and what would be the
appropriate order parameter for them? At this stage, it is important
to distinguish between two types of the quasi-particles in the system.
The first is the aforementioned composite fermions that, as shown in the following,
form Cooper pairs in a similar fashion as electrons in the BCS theory
with exactly the same statistics. The second is vortices that, in a direct
analogy to their counterpart in the $p+ip$ superconductivity, may possess
non-Abelian statistics~\cite{NayakWilczek,Ivanov}. In this study, we are interested
in the superconducting nature of composite fermions.

The superconducting order parameter is conventionally defined in
the BCS theory as the expectation value of two fermion creation
operators, say, $c^{\dagger}_{\bf k}$ and $c^{\dagger}_{-{\bf
k}}$, in the fixed-phase coherent ground state. For fixed-number
systems, an alternative but physically identical formulation of
the BCS superconducting order parameter is given in terms of the
matrix element between the ground states in the $N$ and $N+2$
systems~\cite{Schrieffer}. In this work, we use such formulation
and compute the superconducting order parameter in a numerically
exact manner without any (BCS or otherwise) approximation. Instead
of the usual momentum space representation, it is much more
convenient here to compute the superconducting order parameter in
real space: $F({\bf r})=\langle N+2 | c^{\dagger}({\bf r})
c^{\dagger}({\bf 0}) | N \rangle$ where $| N+2 \rangle$ and $| N
\rangle$ are the ground states in the $N+2$ and $N$ systems,
respectively. The preceding definition, however, is not yet
appropriate for the 5/2 FQHE since the true elementary
quasi-particle is not an electron, but a composite
fermion~\cite{Comment}. A necessary condition for creating a
composite fermion is that the magnetic flux should be increased by
$2\phi_0$, i.~e., $Q \rightarrow Q+1$, at the same time as an
electron is created. For this reason, a correct definition for the
5/2 FQHE superconducting order parameter may be given as follows:
\begin{eqnarray}
F_*({\bf r})=\langle N+2, Q+2 | c^{\dagger}_*({\bf r})
c^{\dagger}_*({\bf 0}) | N, Q \rangle \;, \label{order_parameter}
\end{eqnarray}
where $| N+2, Q+2 \rangle$ and $| N, Q \rangle$ are the ground
states for the particle-flux sector of $(N+2, Q+2)$ and $(N, Q)$,
respectively. In the above, $c^{\dagger}_*$ indicates the
composite fermion creation operator accompanied with an
appropriate increase in flux. We emphasize that
Eq.~(\ref{order_parameter}) is considered an appropriate order
parameter since it is proportional to the superconducting gap
within the BCS theory~\cite{Schrieffer}. It vanishes in the normal
state not by symmetry reasons, but due to gap vanishing.
Finally, it is interesting to note that the composite fermion pair operator
used in Eq.~(\ref{order_parameter}) is conceptually similar to the
composite boson operator considered by Read~\cite{composite_boson_1}
and by Rezayi and Haldane~\cite{composite_boson_2}, which
exhibits Bose-Einstein condensation at odd-denominator fractions.

\begin{figure}[t]
\begin{center}
\mbox{\includegraphics[width=3in]{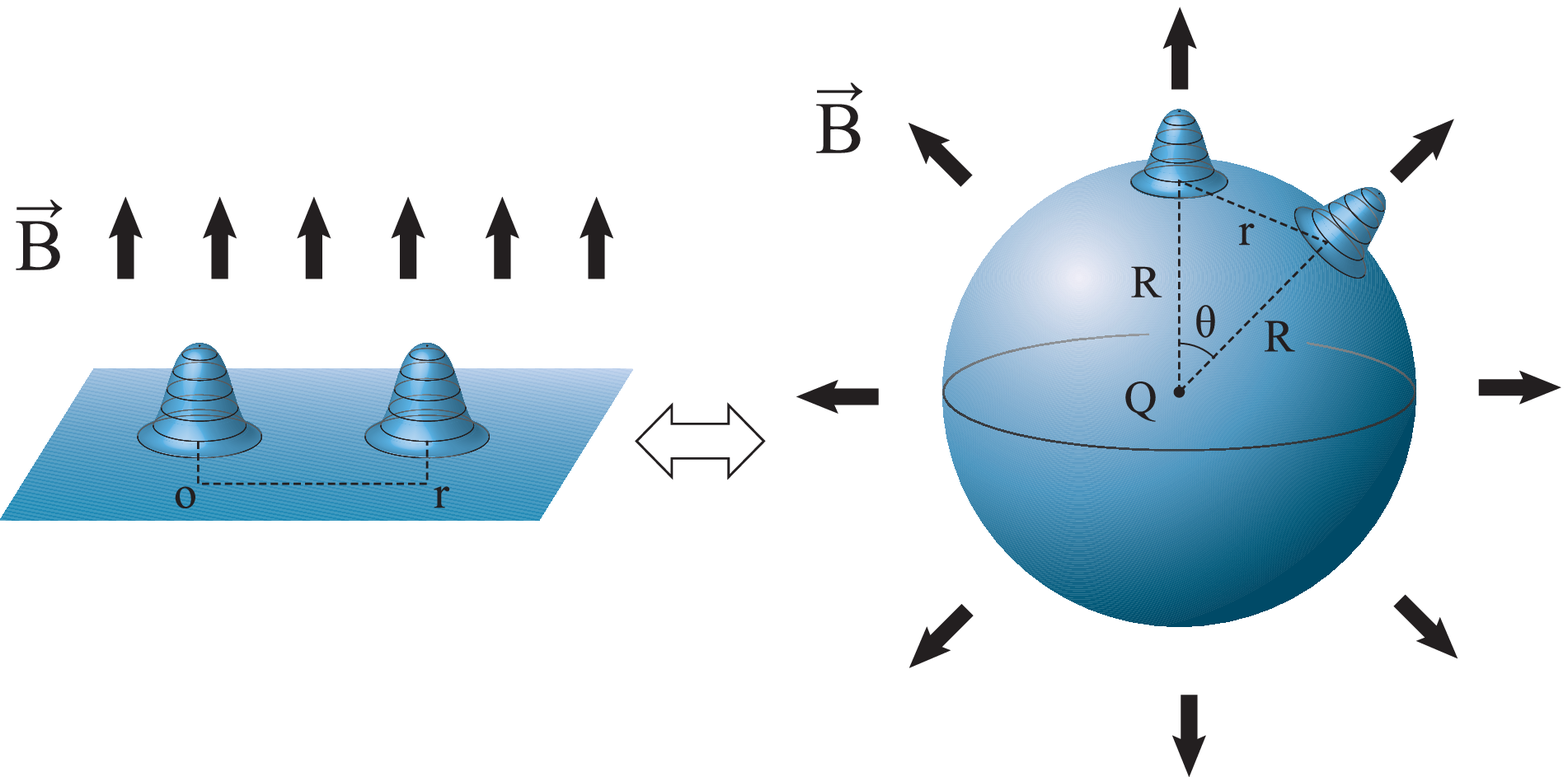}}
\mbox{\includegraphics[width=3in]{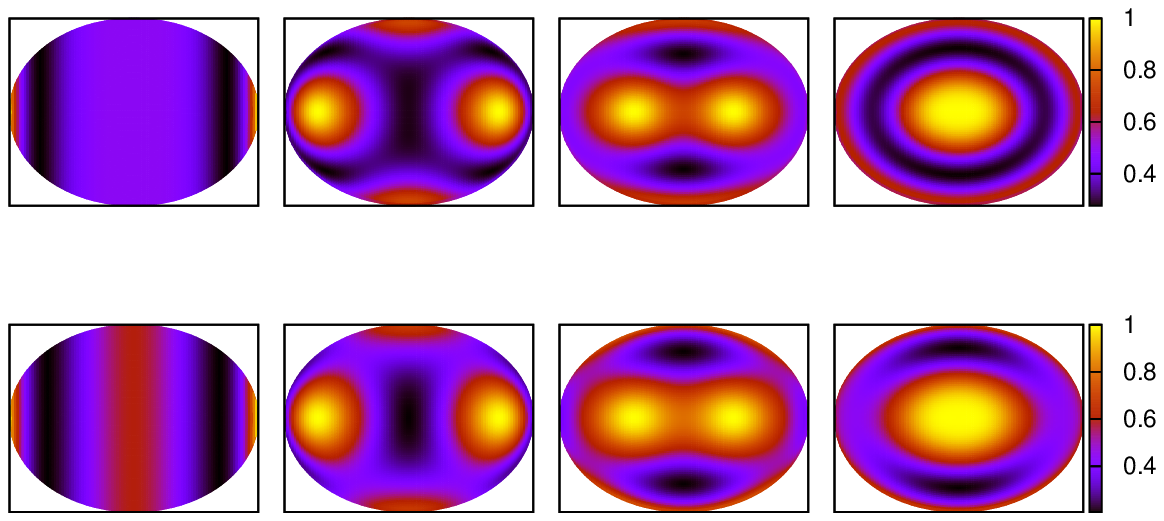}}
\end{center}
\caption{ {\bf Creation of a composite-fermion Cooper pair} {\bf
a}, Schematic diagram for the finite-size spherical geometry used
in our numerical computation. {\bf b}, Resulting density profile
of a composite-fermion Cooper pair as a function of separation
distance in the system of $N=14$ (which includes two electrons
participating in the created Cooper pair). Density profiles are
computed in both cases of $H_2$ (top panels) and the SLL $H_C$
(bottom panels) for comparison. The far left panels depict the
situation where two composite fermions (bright yellow spots) are
most separated, i.~e., one located in the north pole and the other
in the south pole. On the other hand, the far right panels show
that two composite fermions are almost on top of each other. Note
that, with separation distance being measured along the chord,
figures show the correct distance while the shape is distorted.} \label{fig2}
\end{figure}

The composite fermion creation, however, requires more than just
an increase in total flux: two flux quanta should be locally bound
to an electron. In fact, the reason for the apparent absence of
the Meissner effect is that composite fermions do not respond to
external magnetic fields at half filling since all magnetic fields
are captured away in the very process of forming composite
fermions. Recalling that a flux quantum is actually equivalent to
a correlation hole with $2\pi$ phase twist, what one has to do is
to make sure that appropriate correlation holes are bound to each
added electron. Unfortunately, however, it is close to impossible
to derive the CF creation operator as a function of that of
electrons in a closed analytic form since a composite fermion is
an extended object containing full many-body correlations. In this
study, we take a different approach. Since we do not know a priori
how correlation holes are attached to the constituent electron
inside a composite fermion, we let the system evolve to reach the
lowest energy configuration. In other words, we perform exact
diagonalization in the increased particle-flux sector of $(N+2,
Q+2)$ under the constraint that an electron is pinned at the
origin and the other at the position of ${\bf r}$. Let us call the
lowest energy state in this situation $| \psi^{N+2,Q+2}_*({\bf r},
{\bf 0}) \rangle$. Relegating the discussion for details to the
following paragraph, we show the density profile of $|
\psi^{N+2,Q+2}_*({\bf r}, {\bf 0}) \rangle$ in Fig.~\ref{fig2}.

Let us now discuss how to obtain $| \psi^{N+2,Q+2}_*({\bf r}, {\bf
0}) \rangle$. To this end, it is necessary to know the
LLL-projected operator for creating an electron at the position of
${\bf r}$, which is given by:
\begin{eqnarray}
c^{\dagger}_{proj}({\bf r}) = \sqrt{\frac{4\pi}{2Q+1}} \sum_m
\overline{Y}_{QQm}(\theta,\phi) c^{\dagger}_m \;,
\end{eqnarray}
where ${\bf
r}=R(\sin{\theta}\cos{\phi},\sin{\theta}\sin{\phi},\cos{\theta}-1)$
with the radius $R = l_B \sqrt{Q}$ and the magnetic length
$l_B=\sqrt{\hbar c/ eB}$. $\overline{Y}_{Qlm}$ is a complex
conjugate of the monopole harmonics with the monopole strength,
$Q$, the total angular momentum, $l$, and the $z$-component
angular momentum, $m$. $c^{\dagger}_m$ is the usual operator for
creating an electron in the $m$-eigenstate. The LLL-projected
density operator is then defined by $\hat{\rho}_{proj}({\bf
r})=c^{\dagger}_{proj}({\bf r})c_{proj}({\bf r})$ with normalization
that the maximum value of density is unity. Equipped with
the LLL-projected density operator, one can obtain
$|\psi^{N+2,Q+2}_*({\bf r}, {\bf 0}) \rangle$ by using the
Lagrange multiplier method. That is, the constraint of pinning
electrons at the origin and the ${\bf r}$ position are imposed by
introducing the Lagrange multiplier term in the Hamiltonian:
\begin{eqnarray}
H = H_{int} -\lambda_1 [\hat{\rho}_{proj}({\bf r})-1] -\lambda_2
[\hat{\rho}_{proj}({\bf 0})-1] \;,
\end{eqnarray}
where $H_{int}$ is the inter-particle interaction Hamiltonian. The
Lagrange multipliers, $\lambda_1$ and $\lambda_2$, are determined
via imposing the saddle-point conditions for the ground state
energy: $\partial \epsilon_{gr}/\partial \lambda_1 =
\partial \epsilon_{gr}/\partial \lambda_2 = 0$. In the end,
$|\psi^{N+2,Q+2}_*({\bf r}, {\bf 0}) \rangle$ is given as the
ground state satisfying the above saddle-point conditions. It is
worthwhile to mention that, when two electrons are separated
across the north and south poles, the constraint can be imposed
precisely without recourse to the Lagrange multiplier method. We
have verified that the results from the Lagrange multiplier method
are completely consistent with that of the precise method.

\begin{figure}
\begin{center}
\mbox{\includegraphics[width=3in]{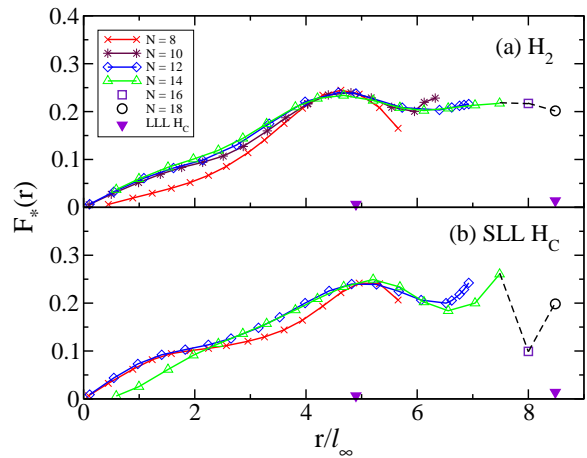}}
\end{center}
\caption{ {\bf Superconducting order parameter as a function of
separation distance between two constituent composite fermions in
a CF Cooper pair} The top and bottom panels show the results for
$H_2$ and the SLL $H_C$, respectively. Here, $N$ indicates the
total number of electrons including the two participating in the
created Cooper pair. Note that, for the $N=16$ and $18$ systems,
the Hilbert space becomes too large to diagonalize except when two
added composite fermions are separated across the north and south
poles. Dashed lines is a guide to eye. In order to accentuate the
distinction between the superconducting FQHE state at $\nu=5/2$
and the usual non-superconducting states, we compute the
superconducting order parameter for the ground state of the
half-filled lowest Landau level, the results of which are denoted
by down triangles in the figure.
Note that only two data points, $N=6$ and $18$, are available for the
half-filled lowest Landau level since the CF sea state is unstable
unless the number of electrons comprising the CF sea (which does not
include two electrons participating in the added Cooper pair) is equal to
that of the filled shell structure, i.~e., $N-2=n^2$ with $n$ being a positive integer.
Also, only even particle numbers are relevant for possible pairing.
} \label{fig3}
\end{figure}

The final ingredient for computing the superconducting order
parameter of the 5/2 FQHE is the realization that
$|\psi^{N+2,Q+2}_*({\bf r}, {\bf 0}) \rangle$ obtained from exact
diagonalization is actually a normalized wave function:
\begin{eqnarray}
| \psi^{N+2,Q+2}_*({\bf r}, {\bf 0}) \rangle =  \frac{
c^{\dagger}_*({\bf r}) c^{\dagger}_*({\bf 0}) | N, Q
\rangle}{\sqrt{\langle N, Q | c_* ({\bf 0}) c_* ({\bf r})
c^{\dagger}_*({\bf r}) c^{\dagger}_*({\bf 0}) | N, Q \rangle} }
\;,
\end{eqnarray}
where normalization is insured by the denominator which is nothing
but the square root of the pair distribution function for
composite fermion. Arguing that the CF pair distribution function
should be similar to the electronic counterpart, $g(r)$, we arrive
at the following conclusion: $c^{\dagger}_* ({\bf r})
c^{\dagger}_* ({\bf 0}) | N, Q \rangle = \sqrt{g(r)}
|\psi^{N+2,Q+2}_* ({\bf r},{\bf 0}) \rangle $. The CF
superconducting order parameter is finally written as follows:
\begin{eqnarray}
F_*({\bf r})= \sqrt{g(r)} \langle N+2, Q+2 | \psi^{N+2,Q+2}_*
({\bf r},{\bf 0}) \rangle \;.
\end{eqnarray}
Figure~\ref{fig3} shows $F_*(r)$ for both $H_2$ and the SLL $H_C$
obtained from exact diagonalization of $N=8-18$ systems, which are
compared with that of the LLL $H_C$. As one can see, for $H_2$,
the superconducting order parameter exhibits a rather smooth
curve, being robust across various finite-size systems while, for
the SLL $H_C$, there are somewhat more fluctuations even though
the overall behavior is similar. The superconducting order
parameter is essentially negligible for the LLL $H_C$, which,
along with the absence of the even-odd effect, is consistent with
the recent conclusion of Storni, Morf, and Das Sarma~\cite{SMDS}.
It is emphasized that all our numerical results taken together
convincingly establish that the real 5/2 FQHE state has underlying
superconductivity. In addition to providing concrete theoretical
evidence for superconductivity, an important lessen from $F_* (r)$
is that the superconducting order parameter is rather long-ranged,
being sizable at least up to ten magnetic lengths. This is
important since the superconducting coherence length sets a
natural length scale for performing coherent quantum operations.

Now, let us make a connection between our findings and previous
results from field theories. The composite fermion (CF) creation
operator defined in our paper creates a bound state between an
electron and a correlation hole under the condition that the
magnetic field is increased by two flux quanta whenever an
additional electron is added. In the lowest Landau level, a
correlation hole is always concomitant with the formation of a
vortex. Since the magnetic field is increased by two flux quanta,
the created bound state is actually equivalent to that between an
electron and a double vortex. Meanwhile, in the composite fermion
Chern-Simons (CS) gauge field theory, it is generally accepted
that binding a double vortex is adiabatically connected to the
attachment of an infinitesimally thin tube of two flux quanta,
i.~e., Chern-Simons flux, into an electron. In this sense, our CF
creation operator essentially plays the same role as creating
CS-flux attached electrons. According to the theory of Halperin,
Lee, and Read~\cite{HLR}, and also Kalmeyer and Zhang~\cite{KZ},
the ground state at $\nu=1/2$ is the Fermi sea state of the
CS-flux attached electrons and therefore, in this state, the
superconducting order parameter must vanish when measured in terms
of the CS-flux attached electrons. On the other hand, the
Moore-Read Pfaffian state, which is the exact ground state of
$H_3$, by construction contains the superconducting order in terms
of the CS-flux attached electrons. It is for the first time in
this work that these expectations are explicitly shown
to be true in realistic Hamiltonians.



{\bf Acknowledgments} This research was supported in part by the
Korea Science and Engineering Foundation grant funded by the Korea
government (MEST) under Quantum Metamaterials Research Center (K.
P.).

{\bf Author Information} Correspondence and requests for materials
should be addressed to K. P. (e-mail: kpark@kias.re.kr).


\begin{thebibliography}{99}
\bibitem{Comment}
We have directly computed the electronic superconducting order
parameter and confirmed that it becomes vanishingly small rather
quickly as the system grows in size.
\end{thebibliography}
\end{document}